\documentclass[11pt]{article}
\usepackage{graphicx}

\newcommand{\BABARPubYear}    {01}

\newcommand{\BABARProcNumber} {08}
\newcommand{\SLACPubNumber} {8791}

\input pubboard/babarsym

\long\def\inst#1{\par\nobreak\kern 4pt\nobreak
    {\it #1}\par\vskip 10pt plus 3pt minus 3pt}

\def\Journal#1#2#3#4{{#1} {\bf #2}, #3 (#4)}

\def\PRL{\em Phys. Rev. Lett.}

\def\ZPC{{\em Z. Phys.} C}
\def\EPJ{{\em Eur. Phys. J.} C}

\def\ra{\rightarrow}

\def\be{\begin{equation}}
\def\ee{\end{equation}}
\def\bea{\begin{eqnarray}}
\def\eea{\end{eqnarray}}
\def\bes{\begin{equation}}
\def\ees{\end{equation}}
\def\beas{\begin{eqnarray*}}
\def\eeas{\end{eqnarray*}}

\def\bu    {B^+}
\def\bz    {B^0}
\def\bzb   {\overline{B}{}^0} 
\def\BzBzb {B^0 \overline{B}{^0}} 
\def\bs    {B^0_s}
\def\bsb   {\overline{B}{}^0_s} 
\def\FourS {\Upsilon{\rm( 4S)}} 

\begin{document}
{\pagestyle{empty}

\begin{flushright}
SLAC-PUB-\SLACPubNumber \\
\babar-PROC-\BABARPubYear/\BABARProcNumber \\
March 27th, 2001 \\
\end{flushright}

\par\vskip 4cm

\begin{center}
\Large \bf B Mixing and Lifetime Measurements with the \Lbabar\ Detector.
\end{center}
\bigskip

\begin{center}
\large 
Concezio Bozzi\\
INFN Sezione di Ferrara \\
Via Paradiso 12, I-44100 Ferrara, Italy \\
(representing the \lbabar\ Collaboration)
\end{center}
\bigskip \bigskip

\begin{center}
\large \bf Abstract
\end{center}
Recent \lbabar\ measurements on lifetime and 
mixing of $B$ mesons are reported. Various techniques are used, ranging 
from the full reconstruction of hadronic $B$ decays, to partial reconstruction 
techniques, and to a totally inclusive approach with dilepton events. The results presented are 
based on a data sample collected by \lbabar\ during the 1999-2000 data taking, and 
should be considered as preliminary.

\vfill
\begin{center}
Contributed to the Proceedings of the 4$^{th}$ International 
Workshop on B Physics and CP Violation (BCP4), 
2/19/2001---2/23/2001, Ise, Japan
\end{center}

\vspace{1.0cm}
\begin{center}
{\em Stanford Linear Accelerator Center, Stanford University, 
Stanford, CA 94309} \\ \vspace{0.1cm}\hrule\vspace{0.1cm}
Work supported in part by Department of Energy contract DE-AC03-76SF00515.
\end{center}
}

\newpage

\section{Introduction}\label{sec:intro}
A precision measurement of the $\BzBzb$ oscillation 
frequency is of great importance since this quantity is sensitive to the 
CKM matrix element $|V_{td}|$ and, in combination with knowledge of the 
$\bs\bsb$\ oscillation frequency, provides a stringent constraint 
on the Unitarity Triangle. 
A precision measurement of $B$ meson lifetimes is also of great importance for the 
understanding of the dynamics involved in heavy quark decays.  
Moreover, in the specific case of the new asymmetric 
B-Factories, both lifetime and time-dependent mixing measurements 
are powerful tools for understanding  
the performance of both the detector and the analysis algorithms, and therefore 
they represent validation analyses for the measurements of time-dependent CP asymmetries.

This paper presents a set of $B$ meson lifetimes and mixing measurements performed with 
the \lbabar\ detector \cite{babarpaper} at the PEP-II $e^+e^-$ asymmetric B-Factory. 
Several techniques are used, ranging from fully exclusive to completely inclusive 
event selections. Exclusive 
measurements generally offer smaller systematic uncertainties with respect to inclusive 
ones, but they also suffer from smaller branching fractions, and consequently 
larger statistical uncertainties. The following sections will detail the three main mixing 
and lifetime measurement procedures followed in \lbabar\, namely the selection of events 
\begin{itemize}
\item where one $B$ meson decay is fully reconstructed in a hadronic flavour eigenstate, or 
\item containing two high momentum leptons from semileptonic $B$ meson decays 
({\it{dilepton events}}), or 
\item where a $B$ decay is reconstructed by using a semi-exclusive technique in either 
an hadronic or a $D^{*} \ell \nu$ final state. 
\end{itemize}
The data sample used was collected by the \lbabar\ detector at the PEP-II asymmetric B-Factory, during 
the years 1999-2000. Unless otherwise specified, the integrated luminosities 
are 20.7 $fb^{-1}$ on the $\FourS$\ peak, and 2.6 $fb^{-1}$ 40 MeV below resonance. All 
results are preliminary. 

\section{Mixing and Lifetimes with fully reconstructed hadronic $B$ decays}
This technique is the most similar to the measurement of the time-dependent CP asymmetry 
also presented at this Conference \cite{stew,sin2bpaper}, the only difference being the 
exclusive reconstruction of a flavour eigenstate instead of a CP eigenstate. 
The main ingredients necessary to perform these time-dependent measurements are: 
\begin{enumerate}
\item to exclusively reconstruct a flavour-eigenstate hadronic final state ($B_{reco}$), 
\item to tag the flavour of the other $B$ meson ($B_{tag}$)\footnote{This step is not needed for lifetime 
measurements.},
\item to measure the time difference $\Delta t$ between the two $B$ meson decays.
\end{enumerate}
Since $B$ production at the $\FourS$ resonance is coherent, tagging the flavour of a $B$ meson 
at its decay time will unambigously determine the flavour of the other $B$ at the same time. 
The distribution of the time difference $\Delta t$ between $B$ meson decays with the opposite  
($h_+$) or the same ($h_-$) flavour is therefore 
\be \label{eqn:sig}
h_\pm = 1/4 \Gamma e^{-\Gamma |\Delta t|}(1 \pm (1-2w)\cos (\Delta m_d \Delta t))
\ee
where $\Gamma = 1/{\tau_{B^0}}$ and $w$ 
is the probability to get a wrong flavour tagging. The resulting same/opposite 
flavour asymmetry will have a pure cosine time dependence. Equation \ref{eqn:sig} has to 
be modified to take into account the detector resolution function and backgrounds. 

\subsection{Exclusive reconstruction of hadronic $B$ decays}
Several hadronic decays are reconstructed\footnote{charge conjugation is always implied in 
the following.}, based on most favourable production rates and low background contamination:
\\

\begin{center}
\begin{tabular}{rlccrl} 
$\bzb$ $\rightarrow$ & $D^{(*)+}\pi^-$    & & & $B^-$   $\rightarrow$ & $D^{(*)0}\pi^-$\\ 
                     & $D^{(*)+}\rho^-$   & & &                       & $J/\psi K^-$   \\ 
                     & $D^{(*)+}a_1^-$    & & &                       & $\psi(2S) K^-$ \\ 
                     & $J/\psi {\overline{K}{}^{*0}}$ & & \\
\end{tabular}
\end{center}

\begin{figure}[!htb]
\center
\includegraphics[width=0.75\textwidth]{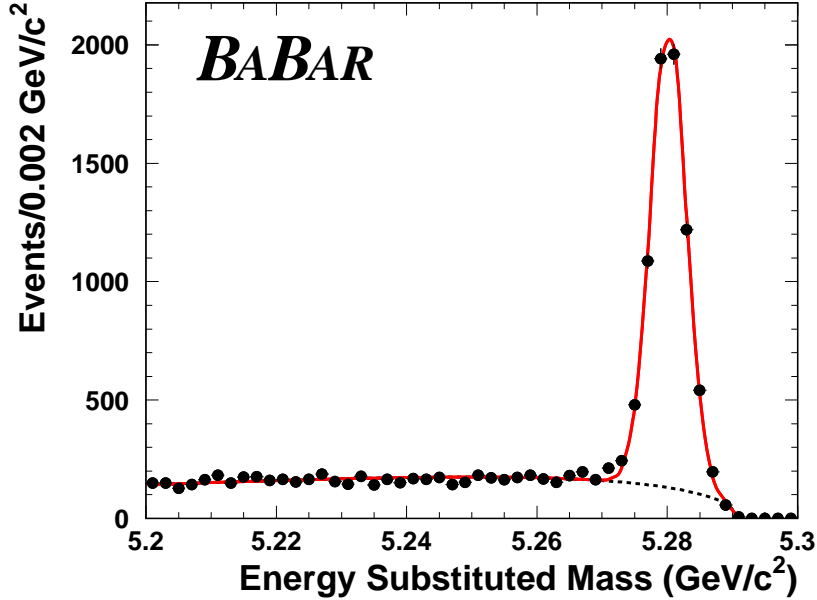}
\caption{Beam energy substituted mass distribution for fully 
reconstructed hadronic neutral $B$ decays.}
\label{fig:m_ES}
\end{figure}

$\,$ \\
\noindent where $D^{*+} \rightarrow D^0\pi^+$, $D^{*0} \rightarrow D^0\pi^0$, 
$\psi(2S) \rightarrow J/\psi \pi^+\pi^-$ or $\ell^+\ell^-$, $J/\psi \rightarrow \ell^+\ell^-$,  
${\overline{K}{}^{*0}} \rightarrow K^-\pi^+$. Neutral $D$ mesons are 
reconstructed in the $K^-\pi^+$, $K^-\pi^+\pi^-\pi^+$, $K^-\pi^+\pi^0$ and 
(except for $B^- \rightarrow D^{*0}\pi^-$) $K_s^0\pi^+\pi^-$ modes, 
charged $D$ mesons in the $K^-\pi^+\pi^+$ and $K_s^0\pi^+$ modes. 
%
%
%
Signal to background discrimination is based on two variables, the 
beam energy substituted mass, $m_{ES}$, and the energy difference $\Delta E$, defined as 
\beas
m_{ES} = \sqrt{E_{beam}^{*2} - p_B^{*2}} & \, \, \, \, \, \, 
{\mathrm{and}} \, \, \, \, \, \, & \Delta E = E_B^*-E_{beam}^*,
\eeas
where $E_{beam}^*$, $p_B^*$ and $E_B^*$ are respectively the beam energy and the momentum and 
energy of the fully reconstructed $B$ meson, computed in the center-of-mass (CMS) frame. 
The $m_{ES}$ resolution is about 2.6 MeV and is dominated by the beam energy spread, whereas 
the resolution on $\Delta E$ is mode-dependent and varies between 20 and 30 MeV, being 
worse in decay chains involving neutral pions. The $m_{ES}$ distribution for neutral $B$ decays, 
after applying a 3$\sigma$ cut on $\Delta E$, is shown in Figure \ref{fig:m_ES}. 
The number of neutral (charged) B signal events with $m_{ES} > 5.27$ GeV/$c^2$ is 6643$\pm$96 
(6928$\pm$94), with a 
purity of about 84\% (87\%). The combinatorial background is estimated by fitting the $m_{ES}$ 
distribution between 5.2 and 5.3 GeV/$c^2$ with the sum of a gaussian for signal 
and an {\it{Argus function}} \cite{argusfunction} for background. This procedure also allows 
to assign a signal probability on an event-by-event basis, which is used in the fitting 
procedure. The time dependence of combinatorial background is determined from the time 
distribution of events in the sideband region. In addition to combinatorial 
background, there is a small contribution which peaks in the signal region. This peaking 
background is due to pion/photon swapping between the two $B$ mesons in the event, and 
introduces a contamination from charged $B$ mesons in neutral $B$ decays and vice-versa. 
Detailed Monte-Carlo studies show that peaking background is at the percent level. 

\subsection{B-flavour tagging and particle identification}
\begin{figure}[!htb]
\center
\includegraphics[width=0.65\textwidth]{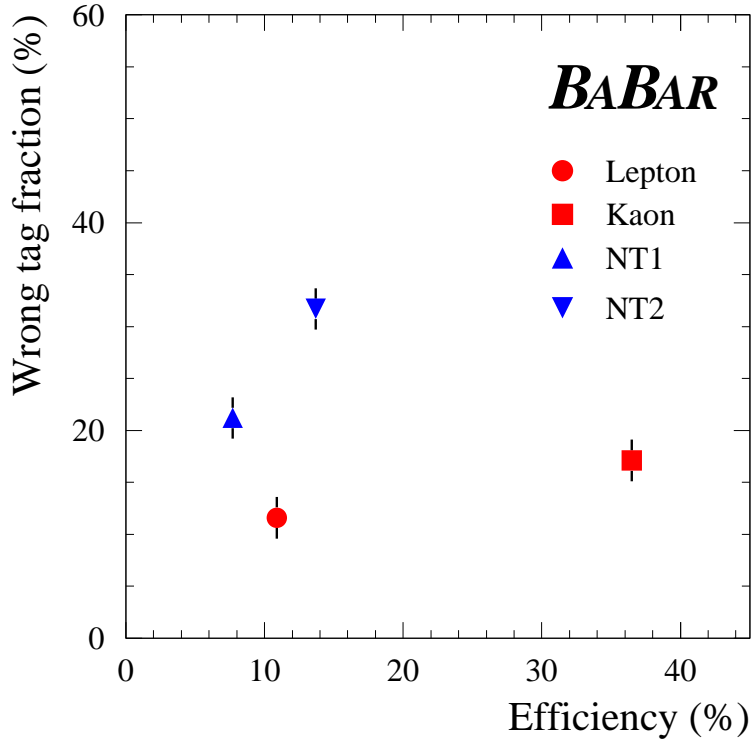}
\caption{Tagging efficiencies vs. wrong tag probability for the four tagging 
categories described in the text.}
\label{fig:tag}
\end{figure}

After reconstructing a full $B$ meson decay chain ($B_{reco}$), 
the remaining tracks of the event are analyzed to determine the $B_{tag}$ flavour. 
Powerful tagging signatures are primary leptons in semileptonic 
$B$ decays, kaons resulting from $b \rightarrow c \rightarrow s$ 
transitions and slow pions from $D^*$ in $B \rightarrow D^*X$ decays. 
The tagging algorithm assigns events to one (and only one) of the following categories, 
in order of decreasing priority: 
\begin{itemize}
\item primary lepton tag (electron with $p>1$ GeV/$c$ in the CMS 
or muon with $p>1.1$ GeV/$c$ in the CMS if no electron is found), with the tag 
determined by the lepton charge;
\item kaon tag, with tagging defined by the sum of all kaon charges (required 
to be non-zero);
\item NT1 and NT2 tags, based on the output of a neural network. 
\end{itemize}
NT1 and NT2 are aimed at recovering leptons which do not pass the requirements for the 
primary lepton tag, resolving ambiguous cases 
where a lepton tag has a conflicting kaon tag, and identifying slow pions from $D^*$ 
decays. Figure \ref{fig:tag} 
shows the tagging efficiency $\varepsilon$ versus the fraction of wrong tags $w$. 
The figure of merit for tagging is defined as $Q=\varepsilon (1-2w)^2$. The kaon tagging 
category has the highest $Q$, followed by lepton, NT1 and NT2 tags. The number 
of tagged neutral $B$ events is 4538$\pm$75, which gives a tagging efficiency of 
68\%. The mistag fractions are determined from the fit to data in the mixing analysis (see \ref{sec:results}).  

Most of the tagging power relies on particle identification \cite{babarpaper}, which has 
therefore to be optimized for maximal efficiency and low misidentification probability. 

{\it{Electron identification}} is based on track matching with a calorimeter cluster, on 
the momentum-to-energy 
ratio (0.89$<E/p<$1.2, where $E$ and $p$ are the energy measured in the calorimeter and the 
track momentum measured in the drift chamber, respectively), on requirements on the 
electro-magnetic shower shape, and on consistency for $dE/dx$ and the  
Cherenkov angle measured in the DIRC (when available) with the electron hypothesis. 

{\it{Muon identification}} relies mainly on the Instrumented Flux Return (IFR), namely on 
the number of measured interaction lengths (greater than 2), on the difference between 
the measured and the expected number of interaction lenghts (less than 1), on 
matching between IFR hits and the extrapolated track, and on requirements on the average 
number and spread of IFR hits per layer. 

{\it{Kaons}} are identified with a neural network based on likelihood ratios computed 
from quantities measured in the DCH and SVT ($dE/dx$), and in DIRC, where single hits are 
compared with the pattern expected for Cherenkov light in the kaon/pion hypotheses. 
Kaon/pion separation is above 3$\sigma$ for momenta up to 3.5 GeV/$c$. 
	
Table \ref{tab:partID} shows the efficiencies and pion misidentification rates for electrons, 
muons and kaons, as determined from data control samples. 
\begin{table}
\begin{center}
\caption{Efficiencies and pion misidentification probabilities for 
electrons, muons and kaons.}\label{tab:partID}
\vspace{0.2cm}
\begin{tabular}{|c|c|c|} 
\hline 
\raisebox{0pt}[12pt][6pt]{Particle} & 
\raisebox{0pt}[12pt][6pt]{Efficiency (\%)} & 
\raisebox{0pt}[12pt][6pt]{Pion misid. (\%)} \\
\hline
\raisebox{0pt}[12pt][6pt]{Electrons} & 
\raisebox{0pt}[12pt][6pt]{91} & 
\raisebox{0pt}[12pt][6pt]{0.13} \\
\raisebox{0pt}[12pt][6pt]{Muons} & 
\raisebox{0pt}[12pt][6pt]{75} & 
\raisebox{0pt}[12pt][6pt]{2.5} \\
\raisebox{0pt}[12pt][6pt]{Kaons} & 
\raisebox{0pt}[12pt][6pt]{85} & 
\raisebox{0pt}[12pt][6pt]{5} \\
\hline
\end{tabular}
\end{center}
\end{table}

\subsection{Vertexing and time measurement}
The time difference between the two $B$ meson decays $\Delta t$ is inferred from the 
distance between their decay vertices along the beam line, $\Delta z$, via the boost factor: 
$$
\Delta t = \Delta z / (\beta \gamma c).
$$
Since $<\beta\gamma>$=0.56 at PEP-II, the average separation between the two $B$ decay vertices 
is about 250 $\mu$m, which is measurable by a vertex detector. The $B_{reco}$ decay vertex  
is conceptually easily identified, whereas the $B_{tag}$ vertex has to be reconstructed 
inclusively by using all other tracks in the event. Particular care has to be taken in order 
to avoid any bias coming from tracks not originating from the tagging $B$ vertex: 
\begin{itemize}
\item neutral long-lived particle ({\it{e.g.}} $K^0$, $\Lambda$) are searched for and used 
in the vertex fit instead of their charged decay products; 
\item tracks originating from secondary (charm) decay vertices are excluded from the vertex 
fit by an algorithm which removes iteratively the track which gives the biggest chi-square 
difference between the vertices reconstructed with and without the track itself. 
The procedure is repeated until the chi-square difference for every remaining track is less than 
6 or until there are no remaining tracks. 
\end{itemize}

Additional kinematic constraints derived from the $B_{reco}$ momentum and decay vertex position, 
from the beam spot position and size and from the knowledge of the 
boost, are used during the reconstruction of the $B_{tag}$ vertex. 

The resulting $\Delta z$ resolution function is parametrized with three gaussians 
(core, tail, outlier) for the mixing fit, and with a sum of a gaussian and the same gaussian 
convoluted with an exponential decay for the lifetime fit. 
Table \ref{tab:reso} reports the resolution parameters for the 
$z$ position of the fully reconstructed vertex and for $\Delta z$. The dominant contribution 
to the $\Delta z$ resolution is due to the tagging vertex. 
\begin{table}
\begin{center}
\caption{Resolution function parameters for the $z$ position of the fully reconstructed 
vertex and for $\Delta z$.}\label{tab:reso}
\vspace{0.2cm}
\begin{tabular}{|c|c|c|} 
\hline 
\raisebox{0pt}[12pt][6pt]{Parameter} & 
\raisebox{0pt}[12pt][6pt]{$z_{Breco}$} & 
\raisebox{0pt}[12pt][6pt]{$\Delta z$} \\
\hline
\raisebox{0pt}[12pt][6pt]{Core $\sigma$($\mu$m)} & 
\raisebox{0pt}[12pt][6pt]{45} & 
\raisebox{0pt}[12pt][6pt]{100} \\
\raisebox{0pt}[12pt][6pt]{Core fraction (\%)} & 
\raisebox{0pt}[12pt][6pt]{80} & 
\raisebox{0pt}[12pt][6pt]{70} \\
\raisebox{0pt}[12pt][6pt]{RMS ($\mu$m)} & 
\raisebox{0pt}[12pt][6pt]{70} & 
\raisebox{0pt}[12pt][6pt]{170} \\
\hline
\end{tabular}
\end{center}
\end{table}
In the mixing fit, the widths of the core and tail gaussians are scaled on an event-by-event 
basis with the error computed from the vertex fits, whereas the outlier gaussian has a 
fixed width of 8$ps$. A possible remaining bias due to secondary vertex tracks not removed in
the fitting procedure is taken into account by allowing a non-zero mean of the core and tail gaussians. 
Since the bias depends on tagging category, different core biases are allowed for each of them. 
Most of the resolution function parameters are fitted in data: two scale factors for the core 
and tail widths, one tail and four core biases, the relative core/tail/outlier amounts, and a bias for the 
outliers.

\subsection{Results} \label{sec:results}
\begin{figure}[!htb]
\center
\includegraphics[width=0.6\textwidth]{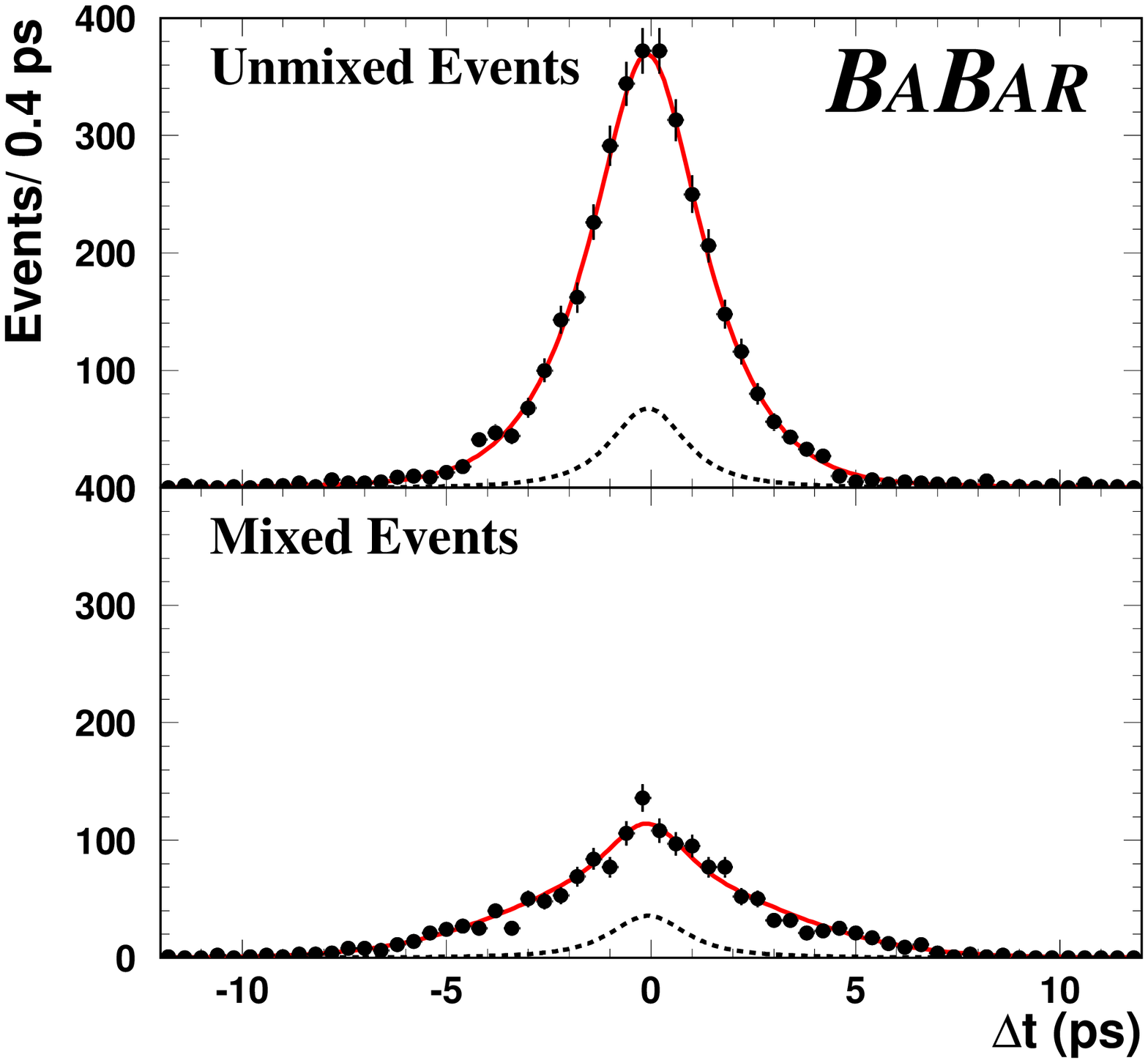}
\includegraphics[width=0.6\textwidth]{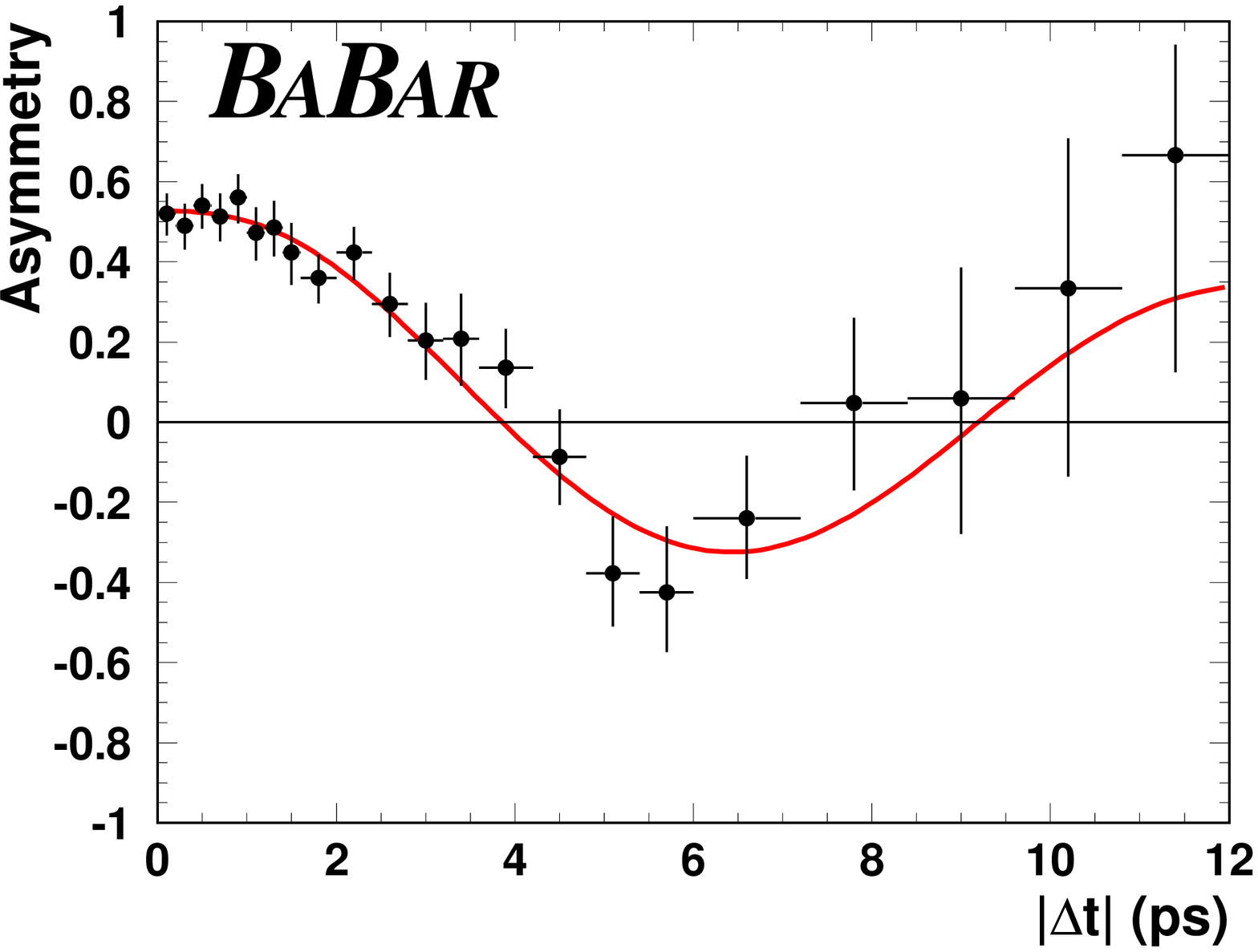}
\caption{Top: $\Delta t$ distributions for unmixed and mixed neutral $B$ events. 
Bottom: time dependent asymmetry. The points represent data, 
the continuous line is the fit result, the dashed line is the background contribution.}
\label{fig:mixunmix}
\end{figure}

An unbinned maximum likelihood technique is applied to fit simultaneously the distribution 
of mixed and unmixed events. The signal is parametrized as in Equation \ref{eqn:sig}, whereas the 
combinatorial background is taken into account by a zero-lifetime and a non-oscillatory, non-zero 
lifetime components, both convoluted with a two-gaussian background resolution function. 
Fit parameters are 34: $\Delta m_d$, the mistag fractions $w$ for the four tagging 
categories, the signal and background resolution functions, and the background parameters. 
The correlation between any other parameter and $\Delta m_d$ is less than 10\%. 
The $\Delta z$ distributions for mixed and unmixed events and the resulting 
time-dependent asymmetry are shown in figure \ref{fig:mixunmix}, together with the fit result. 
The neutral $B$ oscillation frequency is measured to be 
\beas
\Delta m_d = (0.519 \pm 0.020_{stat} \pm 0.016_{syst}) \, \hbar \, ps^{-1}, 
\eeas
and the systematic uncertainty is dominated by resolution and background modeling. 

\begin{figure}[!htb]
\center
\includegraphics[width=0.6\textwidth]{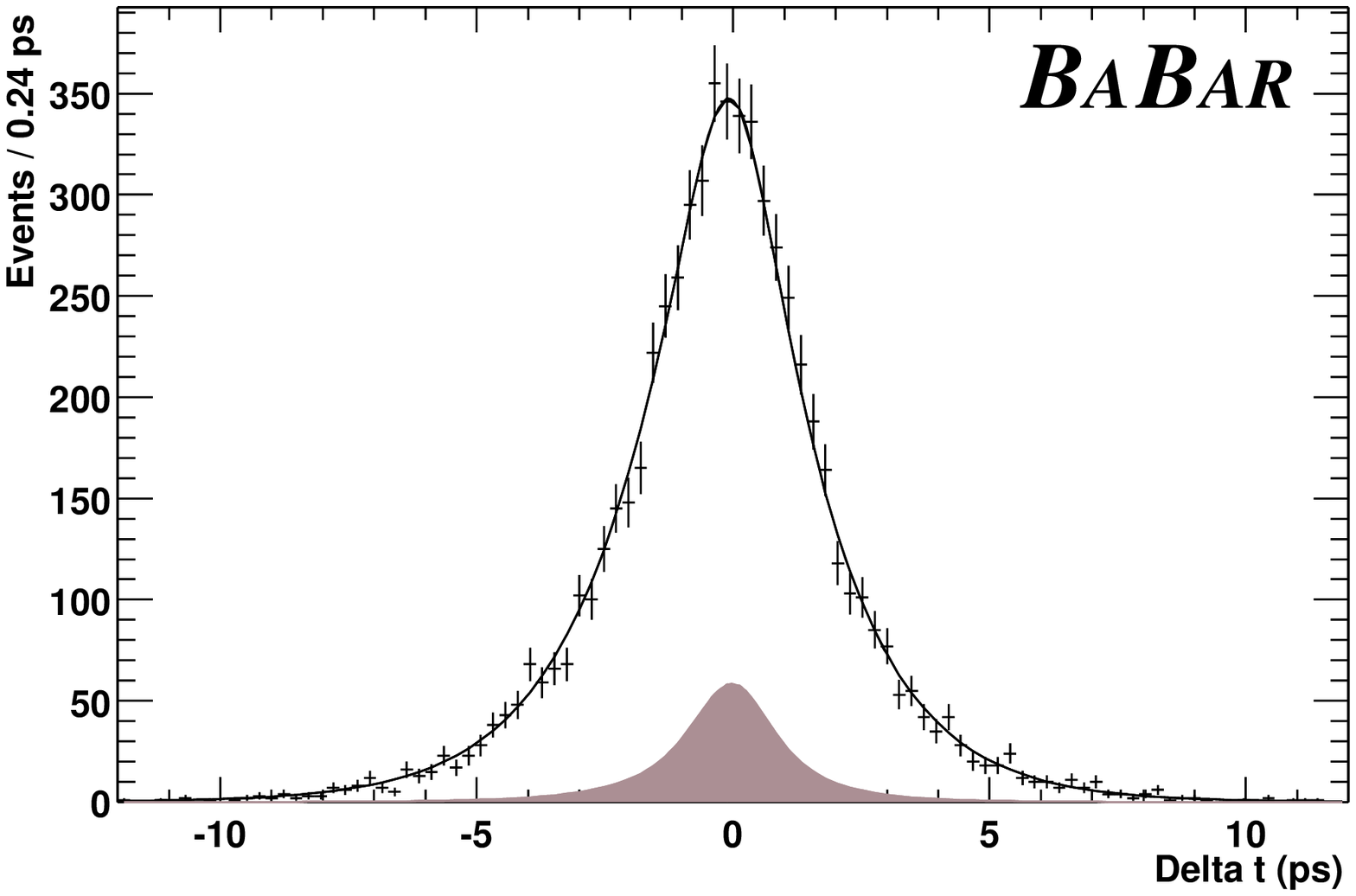}
\includegraphics[width=0.6\textwidth]{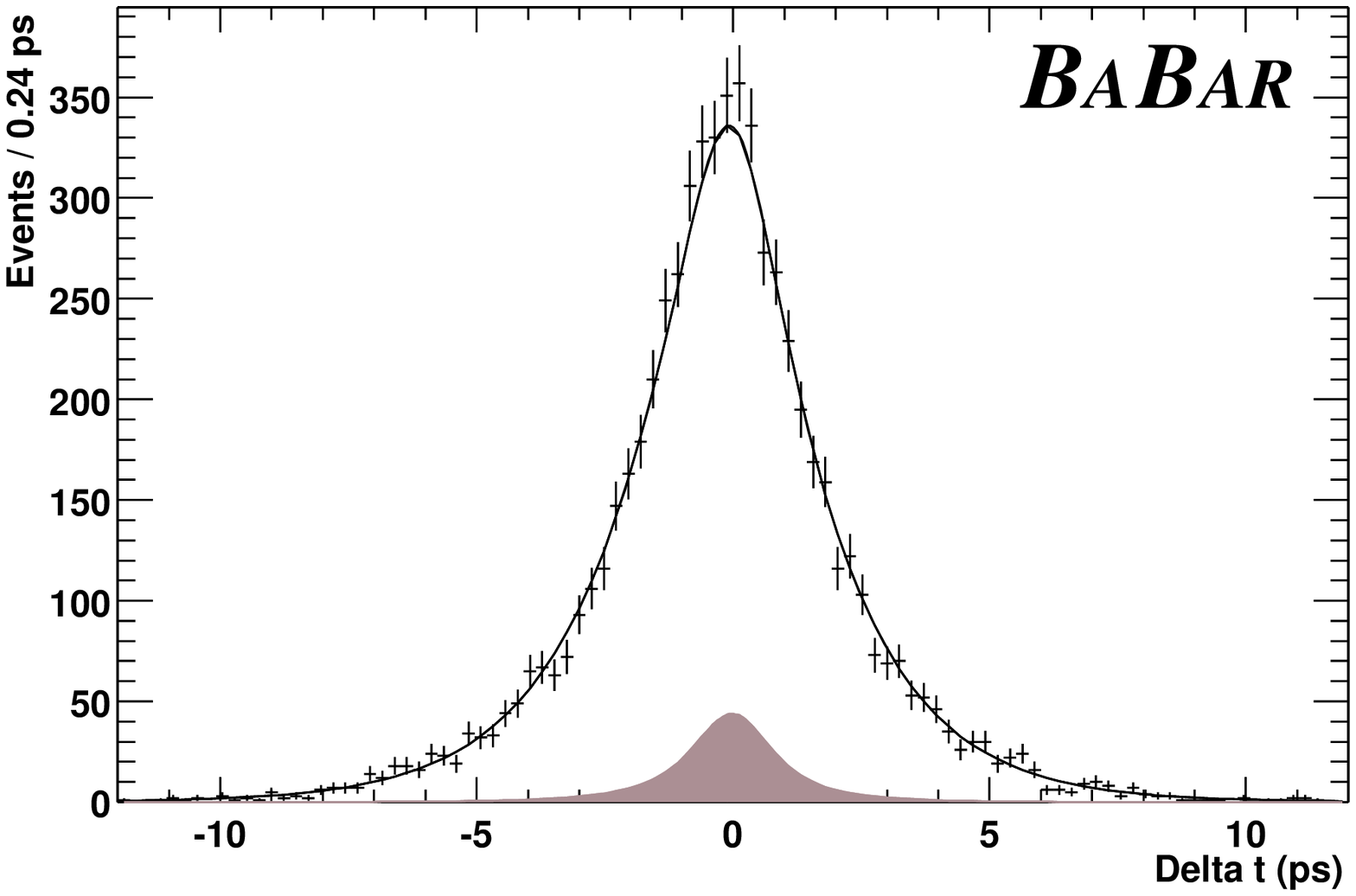}
\caption{$\Delta t$ distributions for neutral (top) and charged (bottom) fully reconstructed 
$B$ candidates. The points refer to data, the line is the fit result, the hatched area represents 
background.}
\label{fig:lifeian}
\end{figure}

Shortly after this Conference, the charged and neutral $B$ meson lifetime measurements 
presented in ICHEP2000 \cite{lifeOsaka} were also updated to the full 1999-2000 data sample 
(figure \ref{fig:lifeian}). 
The new results are 
\beas
\tau_{\bz} & = & (1.546 \pm 0.032_{stat} \pm 0.022_{syst}) \, ps,\\
\tau_{\bu} & = & (1.673 \pm 0.032_{stat} \pm 0.022_{syst}) \, ps,\\
\tau_{\bu}/\tau_{\bz} & = & 1.082_{stat} \pm 0.026 \pm 0.011_{syst},
\eeas
in good agreement with world averages. 

\section{Mixing with dilepton events}\label{sec:dilepmixing}
Measuring $B$ meson properties relying only on two leptons from double semileptonic $B$ decays is 
conceptually simpler than a fully exclusive analysis, and more powerful in statistical terms, 
due to the relatively large semileptonic branching fractions and high lepton identification 
efficiencies. Moreover, the $B$ flavour is easily identified by the charge of the lepton. 
The asymmetry between unlike- and like-sign dilepton events evolves as 
$$\frac{N_{\ell^+\ell^-}(\Delta t) - N_{\ell^+\ell^+,\ell^-\ell^-}(\Delta t)}{
N_{\ell^+\ell^-}(\Delta t) + N_{\ell^+\ell^+,\ell^-\ell^-}(\Delta t)}=
\cos{(\Delta m_d \Delta t)}.$$ 
However, the main drawbacks of this approach are 
the presence of non negligible backgrounds due to leptons 
from the $b\rightarrow c\rightarrow \ell$ decay chain ({\it{cascade decays}}), which are 
also the main source of wrong tags, and most seriously, the presence 
of charged and neutral $B$ mesons in almost equal proportion in the selected sample, which dilutes 
the mixing asymmetry. 

Background from cascade decays is minimized with a neural network technique 
which uses five input variables (the lepton momenta and opening angle, and the total and 
missing energies, all in the CMS), and  
offers better performance in terms of both signal efficiency and mistag rate with respect 
to a traditional approach based on high momentum cuts. 

The fraction $R$ of charged $B$ in the selected sample can be either fit in 
data or taken (with further assumptions) from previous measurements at the $\FourS$. In the 
former case, the statistical accuracy on the mixing measurement decreases substantially, 
due to the high correlation between the oscillation frequency $\Delta m_d$ and $R$. In the latter 
case the systematic uncertainty is dominated by the knowledge of $R$. With the current 
data sample either approach gives similar total uncertainties. However, the present analysis 
fits $R$ because in the fitting procedure this parameter absorbs any difference 
in efficiency or mistag rate between charged and neutral $B$ decays, which would be otherwise another 
source of systematic uncertainty. 

Other selection criteria of the dilepton analysis 
include cuts on event shape variables to suppress continuum background, 
on track quality to improve the $\Delta z$ resolution function, and invariant mass cuts to 
reject backgrounds from $J/\psi$ decays and photon conversions. 
The signal purity 
after all cuts is 78\%. The main backgrounds are 
due to events with at least one lepton from cascade decays (12\%), to events with 
at least one fake lepton (5\%), and to continuum events (5\%).

The $z$ difference between $B$ decay vertices is defined by the $z$ difference of the point 
of closest approach of the leptons to a $\FourS$ vertex in the transverse plane. This vertex 
is determined with a chi-square fit which uses the 
two lepton tracks and a beam spot constraint. The resulting $\Delta z$ resolution, estimated 
from signal Monte-Carlo events, has core and tail widths of about 90 $\mu$m and 200 $\mu$m 
respectively, and 75\% of the events in the 
core. A comparison of the $\Delta z$ distribution for leptons from inclusive $J/\psi$ decays 
shows an agreement within 10\% between data and simulation. 

\begin{figure}[!htb]
\center
\includegraphics[width=0.75\textwidth]{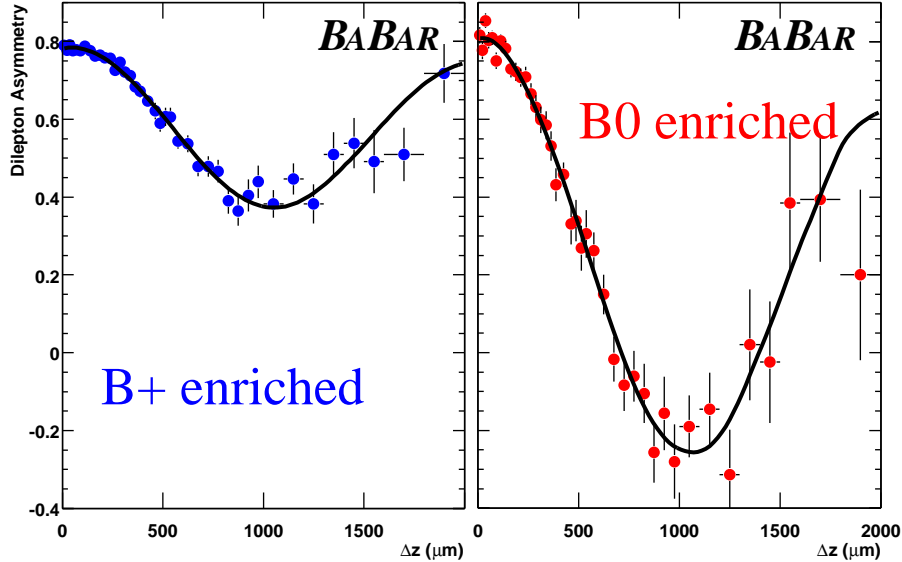}
\caption{$\Delta t$ asymmetry between opposite and same sign dilepton events, for the 
$\bu$-enriched (left) and $\bz$-enriched (right) samples.
The points represent data, the line is the fit result.}
\label{fig:dilepmixing}
\end{figure}

The analysis presented in ICHEP2000 \cite{dilepmixing} has been 
updated shortly after this Conference by using the entire 1999-2000 data sample. About 100000 
events are selected. In the analysis update, the total dilepton sample 
is divided in two independent subsamples, 
enriched respectively in neutral and charged $B$ decays. This is accomplished by reconstructing  
inclusively the $\bzb \rightarrow D^{*+} \ell^- \nu$ decay, where the $D^*$ properties are inferred
from the soft pion produced in its decay \cite{cleo}. Pions of less than 190 MeV/$c$ momentum 
are identified, the $D^*$ energy and momentum are computed and the squared 
missing mass of the neutrino $M_{\nu}^2$ is computed for both lepton/soft pion combinations. 
An event is assigned to the $\bz$-enriched sample if 
a combination satisfies $|M_{\nu}^2| \le 1.5 (GeV/c^2)^2$, to the $\bu$-enriched sample 
otherwise\footnote{if both combinations satisfy the cut,  the one with the smallest 
$|M_{\nu}^2|$ is chosen.}. 
Both subsamples provide similar statistical power to the $\Delta m_d$ measurement. 
A simultaneous binned maximum likelihood fit of the 
same- and opposite-sign leptons time distributions on both subsamples 
gives 
\beas
\Delta m_d = (0.499 \pm 0.010_{stat} \pm 0.012_{syst}) \,\hbar\, ps^{-1}, 
\eeas
which is at present the most precise single measurement of the neutral $B$ meson oscillation frequency. 
Figure \ref{fig:dilepmixing} shows the time-dependent asymmetries for the two subsamples, 
together with the fit results. 
The systematic uncertainties are dominated by the knowledge of the resolution function (0.009), 
the background parametrization (0.006) and the error on the $B$ meson lifetimes (0.004), 
which are fixed to the PDG2000 \cite{PDG2000} values. Separate fits to the two independent subsamples give 
$\Delta m_d = (0.498\pm0.015_{stat}) \,\hbar\, ps^{-1}$ and 
$\Delta m_d = (0.504\pm0.014_{stat}) \,\hbar\, ps^{-1}$ for the 
$\bu$-enriched and $\bz$-enriched samples, respectively. These results are consistent with an independent 
analysis, based on an unbinned maximum likelihood fitting technique, with the same or 
even with a different, momentum-cut based, event selection.

\section{Overview of other lifetimes and mixing results}
The following results were presented at ICHEP2000 \cite{dstarlnumixing,partiallife} 
and are being updated with the full 1999-2000 \lbabar\ data sample. 

\subsection{Mixing from fully reconstructed $\bz \rightarrow D^* \ell \nu$ decays}
This measurement is very similar to the mixing measurement with fully 
reconstructed hadronic decays, the only difference being in the flavour 
eigenstate, namely $\bz \rightarrow D^* \ell \nu$ with $D^* \rightarrow D^0 \pi^+$. 
Neutral $D$ mesons are reconstructed in the $K^- \pi^+$, $K^-\pi^+\pi^-\pi^+$ and $K^-\pi^+\pi^0$ 
decay modes. By using a data sample corresponding to a luminosity of 8.9 $fb^{-1}$ at the 
$\FourS$ resonance, 7517$\pm$104 $\bz \rightarrow D^* \ell \nu$ are reconstructed and the 
mixing parameter is measured to be 
\beas
\Delta m_d = (0.508 \pm 0.020_{stat} \pm 0.022_{syst}) \, \hbar \, ps^{-1}. 
\eeas

\subsection{Lifetimes from partial reconstruction techniques}
Partial reconstruction techniques allow a dramatic enhancement in statistical power, but they are 
more sensitive to the parametrization of physics processes and to biases in the vertex 
reconstruction, the latter due essentially to track swappings between the two $B$ mesons. 
Two decay chains are partially reconstructed: 
\begin{itemize}
\item $\bz \rightarrow D^* \pi$, where the missing mass due to the non-reconstructed $D^0$ meson 
is computed by using only the (fast) pion from the $B$ decay, the (slow) pion from the $D^*$ decay 
and the appropriate kinematic constraints. About 1700 
events are selected by requiring $M_{miss} >$1.854 GeV/$c^2$ in 7.4 $fb^{-1}$ $\FourS$ data. 
\item $\bz \rightarrow D^* \ell \nu$, 
where the same slow pion technique already described in Section 
\ref{sec:dilepmixing} is applied. The signature of a signal event is a pair of oppositely charged 
lepton-soft pion tracks, 
whereas same sign pairs are used to determine the amount (about 40\% of the entire sample) 
and time dependence of the combinatorial background. The analysis selects 
about 90000 signal events on 7.4 $fb^{-1}$ $\FourS$ data.
\end{itemize}
The neutral $B$ lifetime is measured to be 
\beas
\tau_{\bz} & = & (1.55 \pm 0.05_{stat} \pm 0.07_{syst}) \, ps \, \, \, \, (\bz \rightarrow D^* \pi),\\
\tau_{\bz} & = & (1.63 \pm 0.02_{stat} \pm 0.09_{syst}) \, ps \, \, \, \,(\bz \rightarrow D^* \ell \nu).
\eeas

\section{Conclusion}
Asymmetric B-Factories offer the unique possibility of precision lifetimes and mixing measurements in the 
$B$ meson sector. Any single measurement performed with the presently available data is comparable 
to the corresponding 
world average. The \lbabar\ experiment at PEP-II has investigated different reconstruction techniques, 
ranging from the full reconstruction of a $B$ meson decay chain to the totally inclusive selection of 
events with two leptons from semileptonic decays, to the partial reconstruction of hadronic and 
semileptonic decays. Recent preliminary measurements of the charged and neutral lifetimes and 
oscillation frequency with exclusive hadronic decays give 
\beas
\tau_{\bz} & = & (1.546 \pm 0.032_{stat} \pm 0.022_{syst}) \, ps, \\
\tau_{\bu} & = & (1.673 \pm 0.032_{stat} \pm 0.022_{syst}) \, ps, \\
\tau_{\bu}/\tau_{\bz} & = & 1.082 \pm 0.026_{stat}\pm 0.011_{syst}, \\
\Delta m_d & = & (0.519 \pm 0.020_{stat} \pm 0.016_{syst}) \, \hbar \, ps^{-1}. 
\eeas

A preliminary measurement of the mixing parameter with inclusive dilepton events gives
\beas
\Delta m_d = (0.499 \pm 0.010_{stat} \pm 0.012_{syst}) \, \hbar \, ps^{-1}, 
\eeas
which is at present the most precise single measurement of this quantity. 

Further results on lifetimes and mixing with other techniques are being released in a short time scale. 

\section*{Acknowledgments}
I wish to thank my \lbabar\ colleagues Riccardo Faccini, David Kirkby, S\"oren Prell, Gerhard Raven, 
Jan Stark, Vivek Sharma 
and Christophe Y\`eche for useful discussions.

\end{document}